\documentclass[aps,twocolumn,pra,superscriptaddress]{revtex4}
\usepackage{graphicx}
\usepackage{amssymb,amsmath}
\usepackage{moreverb}
\usepackage{color}

\newcommand {\be}{\begin{equation}}
\newcommand {\ee}{\end{equation}}
\newcommand {\ba}{\begin{eqnarray}}
\newcommand {\ea}{\end{eqnarray}}

\begin{document}

\title{Symmetry breaking and singularity structure in Bose-Einstein condensates}
\author{K. A. Commeford}
\affiliation{Department of Physics, Colorado School of Mines, Golden, CO 80401, USA}
\author{M. A. Garcia-March}
\affiliation{Department of Physics, Colorado School of Mines, Golden, CO 80401, USA}
\affiliation{Department of Physics, University College Cork, Cork, Ireland}
\author{A. Ferrando}
\affiliation{Department of Physics, Colorado School of Mines, Golden, CO 80401, USA}
\affiliation{Department d'\'Optica. Universitat de Val\'encia, Dr. Moliner, 50, E-46100 Burjassot (Val\'encia), Spain}
\author{Lincoln D. Carr}
\affiliation{Department of Physics, Colorado School of Mines, Golden, CO 80401, USA}
\affiliation{Physikalisches Institut, Universit\"at Heidelberg, Philosophenweg 12, 69120 Heidelberg, Germany}

\begin{abstract}
We determine the trajectories of vortex singularities that arise after a single vortex is broken by a discretely symmetric impulse in the context of Bose-Einstein condensates in a harmonic trap.
The dynamics of these singularities are analyzed to determine the form of the imprinted motion. We find that the symmetry-breaking process introduces two effective forces: a repulsive harmonic force that causes the daughter trajectories to be ejected from the parent singularity, and a Magnus force that introduces a torque about the axis of symmetry. For the analytical non-interacting case we find that the parent singularity is reconstructed from the daughter singularities after one period of the trapping frequency. The interactions between singularities in the weakly interacting system do not allow the parent vortex to be reconstructed. Analytic trajectories were compared to the actual minima of the wavefunction, showing less than \(0.5\%\) error for small impulse strengths over the entire trajectory.  We show that these solutions are valid within the impulse regime for various impulse strengths using numerical integration of the Gross-Pitaevskii equation. We also show that the actual duration of the symmetry breaking potential does not significantly change the dynamics of the system as long as the impulse strength is small. 
\end{abstract}

\maketitle

\section{Introduction}
\label{sec:introduction}

Vortices are ubiquitous to many diverse branches of science, such as fluid dynamics~\cite{ref:Saffman1992}, meteorology~\cite{ref:DaviesJones2001}, cosmology~\cite{ref:Tanga2002, ref:Barow2007}, liquid crystals~\cite{ref:Pismen1999}, superconductivity~\cite{ref:Essmann1967,ref:Tiley1990, ref:Blatter1994}, solid state physics~\cite{ref:Kosterlitz1973}, and nonlinear singular optics~\cite{ref:Desyatnikov2005, ref:Soskin1998}.  
Vortices have been realized experimentally in Bose-Einstein condensates (BEC), obtained when bosons are cooled down to almost zero temperatures~\cite{ref:Matthews1999, ref:Madison2000, ref:Chevy2000, ref:Anderson2001, ref:AboShaeer2001, ref:Dutton2001, ref:Inouye2001, ref:Raman2001, ref:Leanhardt2002, ref:Shin2004, ref:Scherer2007}. These vortices are expected to offer interesting applications in interferometry~\cite{ref:Thanvanthri2009}, and as a means to study the behavior of random polynomial roots~\cite{ref:Castin2006}.
Vortices are characterized by the presence of a phase singularity to which an integer number can be  associated, called vorticity, topological charge, or winding number~\cite{ref:Berry2007,ref:GarciaMarchPRA2009}. This singularity  behaves as an individual physical entity whose motion can be studied separate from the condensate. The determination of this motion and its control is applicable in the study of many of the fields described above. Here, we determine the dynamics of these singularities when a highly charged vortex in a Bose-Einstein condensate is struck by a symmetry-breaking impulse.  

\begin{figure}[t]
\begin{center}
\begin{tabular}{cccc}
 \hspace{-0.25cm}\includegraphics[width=2cm]{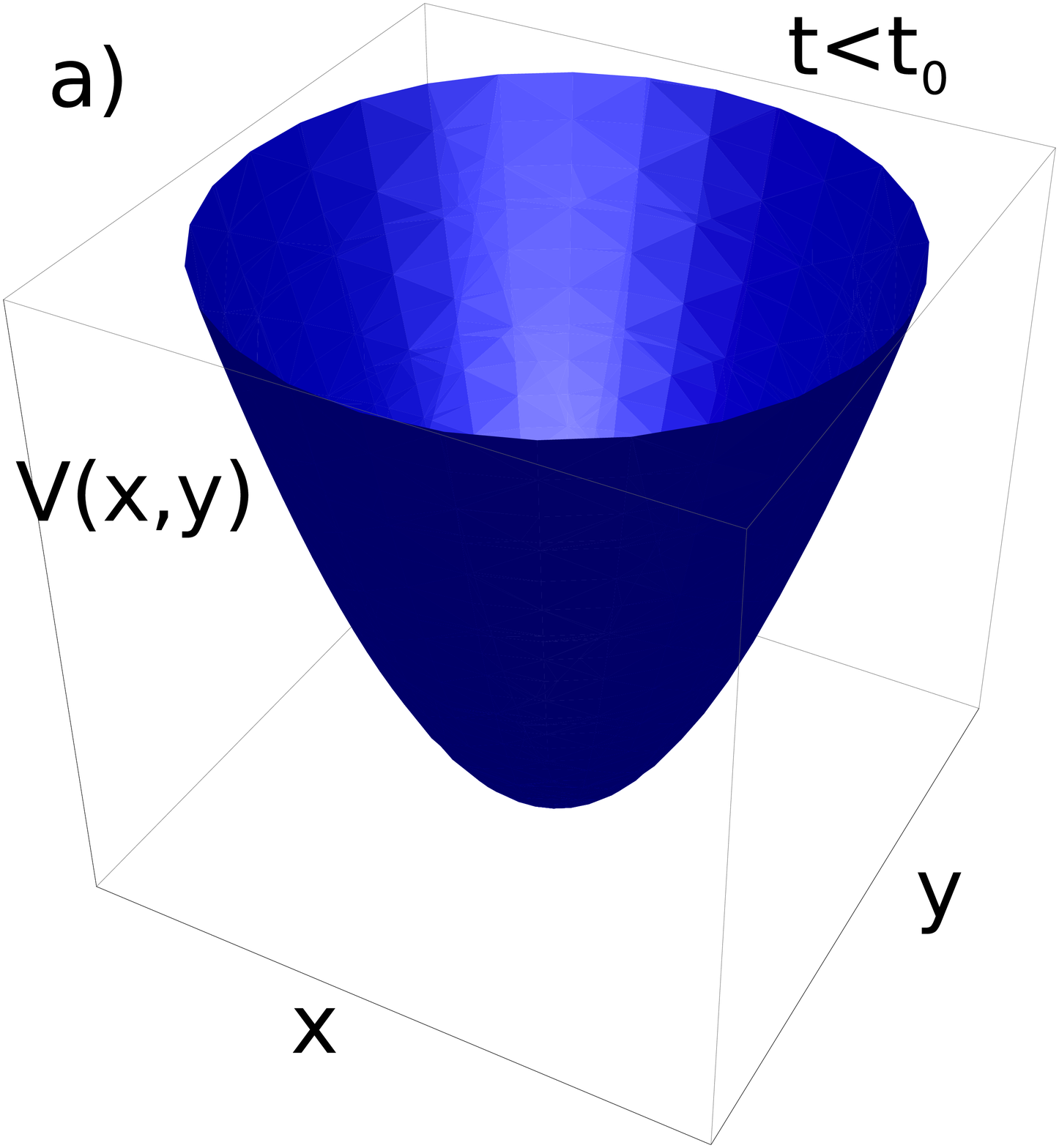}&\hspace{-0.6cm}\includegraphics[width=2cm]{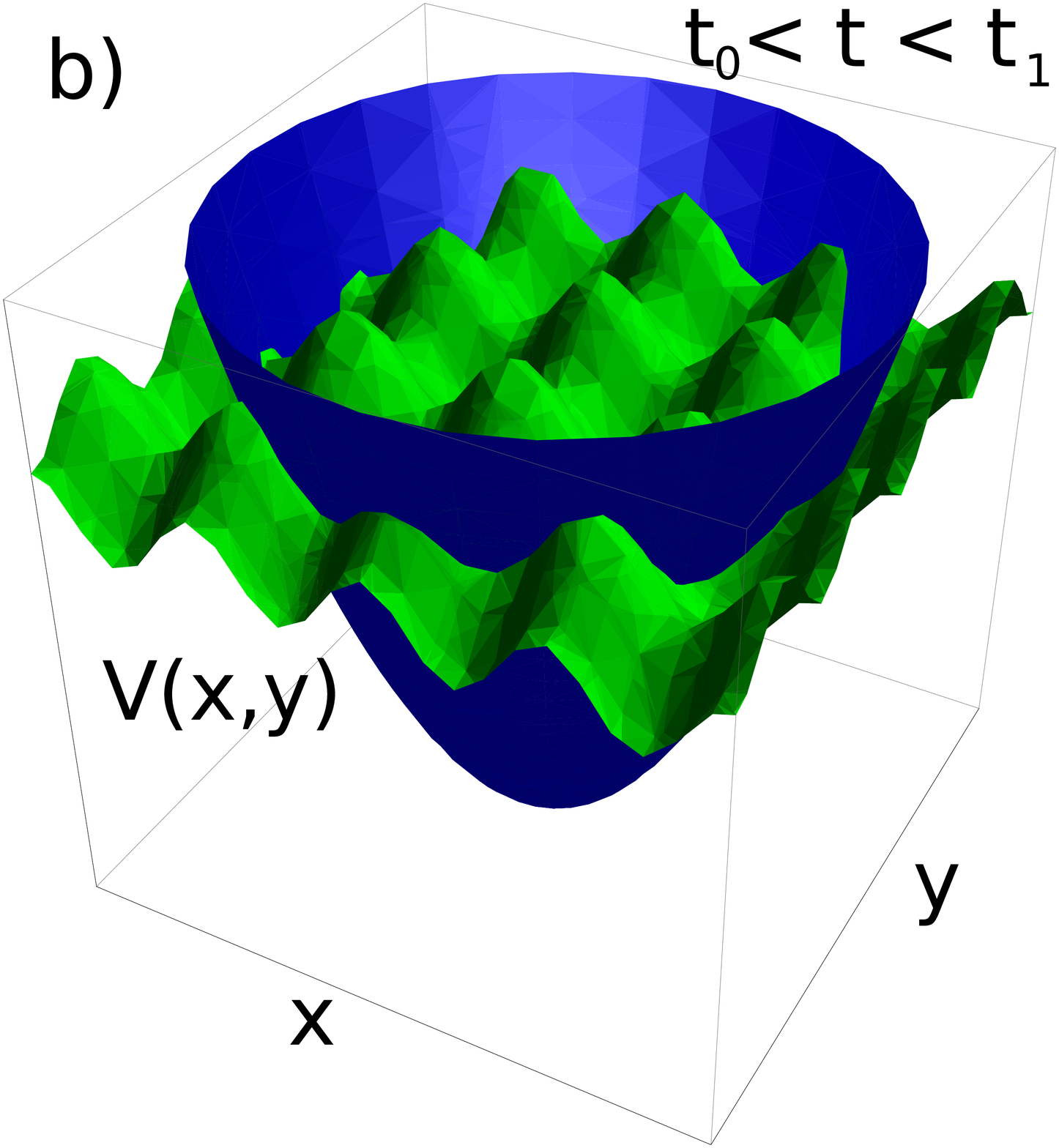}\hspace{-0.6cm}
&\includegraphics[width=2cm]{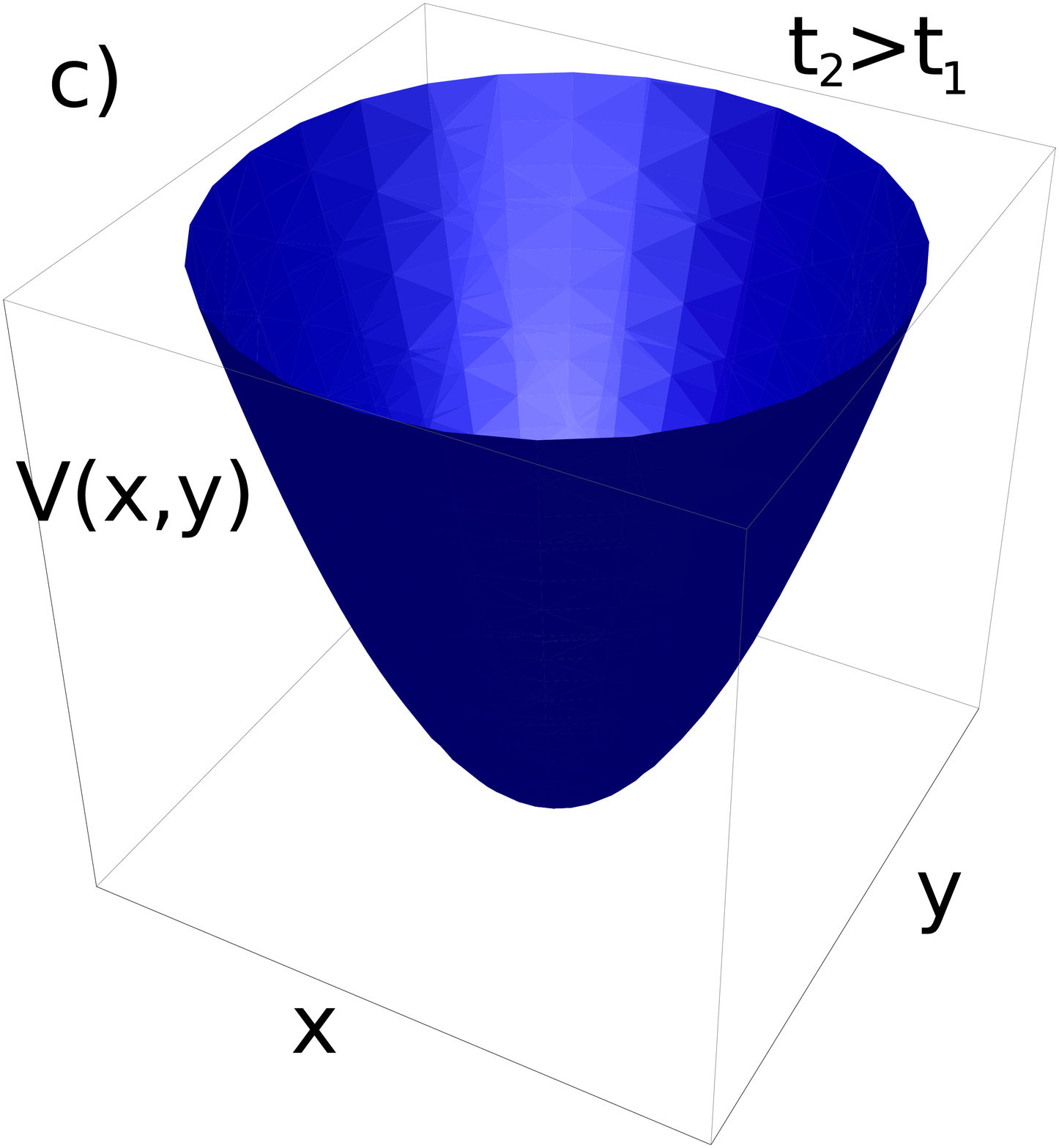}&\hspace{-0.6cm}\includegraphics[width=2cm]{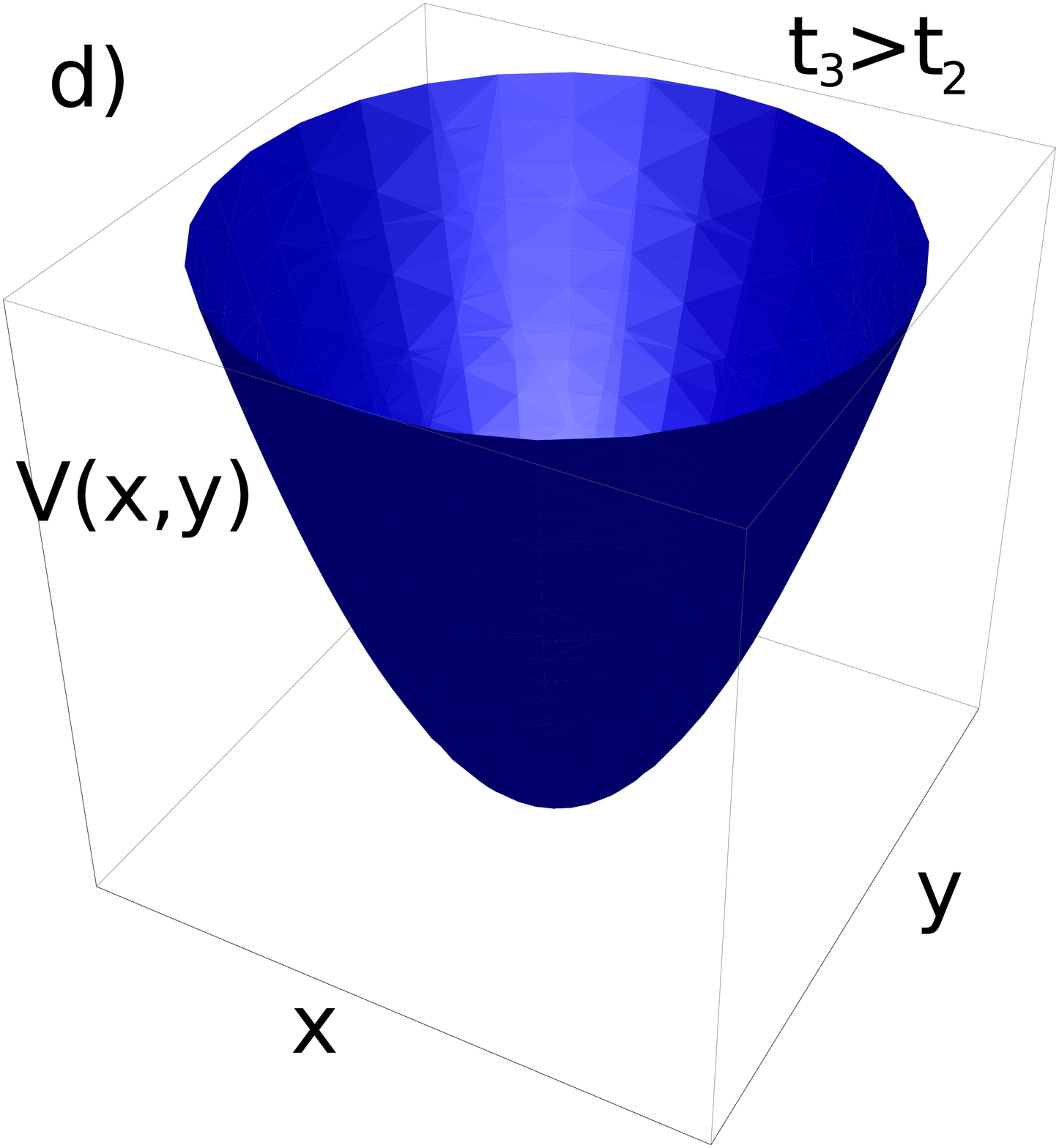}\\
 \hspace{-0.45cm}\includegraphics[width=2.6cm]{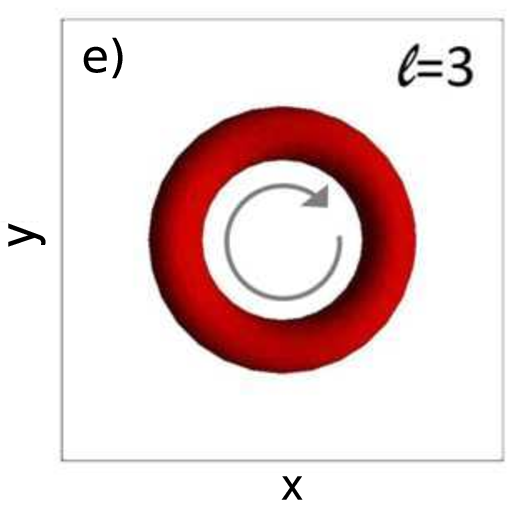}&\hspace{-0.4cm}\includegraphics[width=2.5cm]{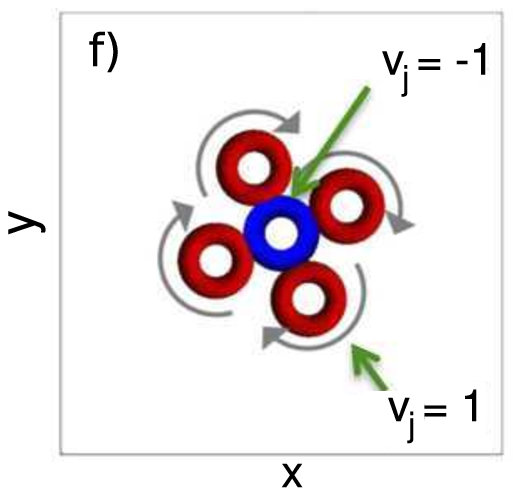}\hspace{-0.4cm}
&\includegraphics[width=2.6cm]{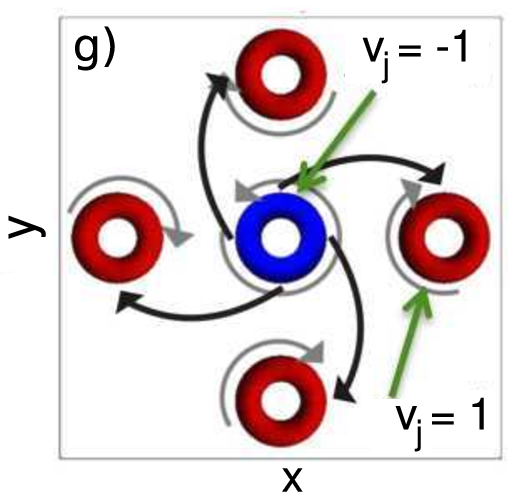}&\hspace{-0.4cm}\includegraphics[width=2.6cm]{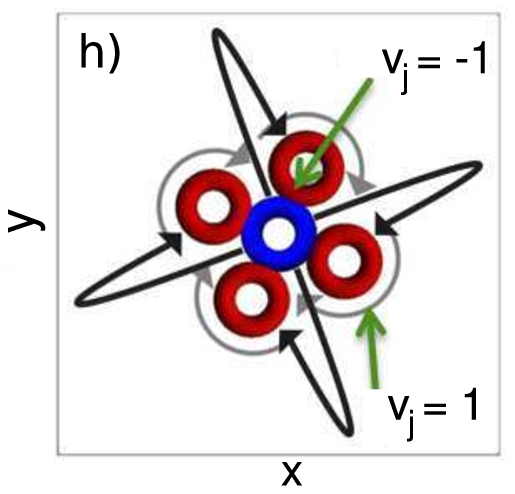}\\
\end{tabular}
\caption{  \label{fig:Fig1}  (Color online). \emph{Schematic of transformation process. } The upper panels represent the potential along time while the lower ones are the corresponding phase singularity structure. The harmonic trap is represented in blue and the symmetry-breaking impulse  with a green surface. The  initial highly charged vortex, represented in e) as a red surface, is broken in five daughter singularities, one at the origin with charge $-1$  (blue surface) and four off-axis of charge $+1$ (red surfaces).  The four off-axis singularities move outwards, eventually coming back close to the origin due to the trap  [black arrows in g) and h)].}
\end{center}
\end{figure}

Specifically, we consider a dynamical situation in which a highly charged two-dimensional vortex is generated in an axisymmetric harmonic potential, and a symmetry-breaking potential is turned on transversely, i.e., in  the plane containing the vortex, for a very short period of time, such that it can be described by an impulse with a potential \(V\) over a period of time \(\Delta t\). This potential shows some rotational discrete point symmetry of order $N$, that is, it reproduces itself under multiple integer rotations of $2\pi/N$~\cite{ref:Hamermesh1964}. The topological charge of the vortex will experience  a transformation, as discussed in~\cite{ref:Ferrando2005PRL, ref:PerezGarcia2007, ref:GarciaMarchPHYSD2009}. We will show that the highly charged parent singularity will disintegrate into a number of single-charged daughter singularities of different sign. The number and sign of these daughter singularities are related to the peculiarities of the symmetry-breaking impulse~\cite{ref:Zacares2009}. Figure \ref{fig:Fig1} shows a representation of this discretely symmetric impulse acting on the parent singularity.

Here we determine the path followed in a harmonic potential by the ejected daughter singularities after the impulse. We obtain these trajectories analytically for the non-interacting case by utilizing the Feynman propagator for a harmonic potential, and determine their validity in the weakly interacting case. In the non-interacting case we find that the parent singularity reconstructs itself from the daughter singularities after a period of time, i.e., there is a full quantum revival.In the interacting case quantum revival is blocked in the mean field theory: repulsion prevents the highly charged parent from being reconstructed, hence describing a helical trajectory around the origin. We discover that there are two effective forces the singularities experience during symmetry breaking; a repulsive harmonic force that causes the trajectories to propagate outward, and a Magnus force that introduces a torque about the axis of symmetry.
Our results pave the way to the control and manipulation of the motion of singularities by means of symmetry-breaking impulses. The results are equally applicable to the neighboring field of nonlinear singular optics~\cite{ref:Soskin1998} by exchanging time evolution with axial-spatial evolution and the symmetry-breaking impulse with a  inhomogeneous thin diffracting  element. 

The study of dynamics of singularities and their interaction is an exciting field with many potential applications. The dynamics of vortex dipoles; their interaction, oscillation, tunneling, and their collapse; has been theoretically studied in the framework of BEC~\cite{ref:Fetter2001,ref:Martikainen2001,ref:Coddington2003,ref:Kevrekidis2003,ref:Watanabe2007,ref:McEndoo2010a,ref:Freilich2010,ref:Neely2010,ref:Kuopanportti2011}. Other structures of singularities and the interactions among them lead to elaborated trajectories \cite{ref:Klein2007, ref:Jackson1999}, as discussed numerically in~\cite{ref:Zhang2010}. The geometry of vortex trajectories, like loops or hyperbolas, is related to vortex creation and annihilation and vortex interactions, and its study leads to a variety of vortex structures~\cite{ref:Bialynicki2000,ref:Bialynicki2001,ref:Berry2001,ref:Infeld2003,ref:Berry2007}. Moreover,  the interpretation of the role of a phase singularity in quantum dynamics is an interesting issue, as well as the effect of the dynamics of the singularities in the quantum system~\cite{ref:Wisniacki2006,ref:Wisniacki2007}.  Also, vortices in BECs, called vortex solitons in nonlinear optics, can show more than one off-axis singularity; in the latter case they are called vortex clusters.~\cite{ref:Crasovan2002,ref:Kobayashi2002,ref:Crasovan2003,ref:Zhou2004,ref:Liu2006,ref:Chen2002,ref:GarciaMarchPRA2009,ref:Geurts2008}. These structures are typically unstable, showing very slow dynamical decay rates, though some controversy has been built up around this issue~\cite{ref:Mottonen2005, ref:Mihalache2006, ref:Pietila2006}. Here, we obtain a breathing cluster of vortices, or planetary vortices, and we obtain numerical stability of this structure along the times of the evolution. 

In Sec. \ref{system}, we define the system and the impulse, and briefly explain the transformation rule used to determine the symmetry of the post-impulse singularity structure.
In Sec. \ref{sec:propagation}, we introduce the initial highly charged vortex, with winding number \(\ell=3\), that will be used in this research as our main test case, for brevity; similar behavior is seen for other initial winding numbers. This section also carries out the harmonic oscillator propagation integral used to determine the wavefunction after the impulse.
Section \ref{sec:trajectories} uses the wavefunction after symmetry breaking to analytically derive the trajectories for the off-axis singularities that are broken out of the initial highly charged vortex by the impulse.
Section \ref{sec:eom} utilizes the trajectories to arrive at analytic descriptions of the equations of motion. These equations of motion are analyzed to understand the fundamental motion that the vortices undergo once symmetry is broken. 
Section \ref{convergence} includes a comparison of the analytic trajectories with the local minima of the wavefunction. The analytic trajectories are also compared to numerical analysis of the Gross-Pitaevskii equation for various values of the nonlinearity, \(g\). In Sec. \ref{sec:conclusion}, we conclude.

\section{Vortices in a Weakly Interacting Bose-Einstein Condensate}\label{system}

Let us consider a system of weakly interacting bosons of mass $M$ confined in a harmonic trap and condensed in the ground state at \(T\ll T_{\mathrm{BEC}}\), thus forming a BEC. We assume that one of the trapping frequencies is sufficiently high to reduce the dimensionality of the system to only two dimensions, but not near any potential resonances. This system can be described by the Gross-Pitaevskii equation (GPE)
\begin{equation}\label{Schrodinger}
i \hbar \partial_t \psi(\mathbf{\tilde{x}},\:t)=H\psi(\mathbf{\tilde{x}},\:t),
\end{equation}
with
\begin{equation}\label{Gross}
H=-\frac{\hbar^2}{2 M}\nabla^2+V(\mathbf{\tilde{x}},t)+g_{\mathrm{3D}}|\psi|^2,
\end{equation}
where $\mathbf{\tilde{x}}=(\tilde{x},\;\tilde{y})\in\mathbb{R}^2$,  and $g_{\mathrm{3D}}$ is the coupling constant, or nonlinearity, defined by \(g_{\mathrm{3D}} = 4 \pi \hbar^2 a_s / M\), where \(a_s\) is the scattering length and \(M\) is the reduced mass, and related to the effective interactions among the bosons in the trap. To model the symmetry breaking impulse, we consider a time-dependent potential given by
\begin{equation}\label{profile}
V({\bf \tilde{x}},\:t)=\left\{ \begin{array}{ll}
V_0({\bf \tilde{x}})& 0\le t<t_0\\
V_0({\bf \tilde{x}})+\Delta V_0({\bf \tilde{x}})&t_0\le t<t_1=t_0 +\Delta t\\
V_1({\bf \tilde{x}})&t_1\le t.\end{array}\right.
\end{equation}
We represent this potential in the upper panel of Fig.~\ref{fig:Fig1}. We assume the length of the second region to be small, \(\Delta t\ll 1\), to model a Dirac delta impulse with constant area but short duration. 
Also, we assume  that \(\Delta V_0\) is invariant under the action of the elements of a discrete rotational group \(C_N\) in order to view the effects of a symmetry-breaking impulse. The original and final media own perfect rotational symmetry, denoted \(O(2)\), with a potential given by 
\begin{equation}
V_0=\frac{1}{2} M \omega^2(\tilde{x}^2+\tilde{y}^2).
\end{equation}
 Mathematically, we express the invariance property of the impulse as
\begin{equation}
\Delta V_0(G{\bf \tilde{x}})=\Delta V_0({\bf \tilde{x}}) \quad \forall G \in C_N.
\end{equation}
Let us consider that for $t< t_0$ the atoms are condensed in a vortex of vorticity $v$, where $v= 1/2\pi \oint_{\Gamma}\nabla  \theta\cdot d\mathbf{l} $ where $\Gamma$ is a closed path encircling the axis of cylindrical symmetry of the vortex~\cite{ref:GarciaMarchPRA2009}. We assume that the phase singularity located at this axis is highly charged, $v>2$. This vortex shows angular momentum  $\ell$ equal to $v$, which is conserved if the symmetry is not externally broken~\cite{ref:GarciaMarchPRA2009}. 

To analyze the effect of discrete symmetry potentials of order $N$ in the properties of vortices,  a quantity called {\it angular pseudomomentum} was associated to them~\cite{ref:FerrandoPRE2006}. To define this quantity, it should be noticed that any vortex solution of the GPE can be written as \(\psi(r,\theta)=e^{i m \theta}u(r,\theta)\), where \(u(r,\theta)=u(r,\theta +\frac{2\pi}{N})\), with $m$ an integer. The effect of such a rotation is the adding of $m$ times the same angle to its phase~\cite{ref:FerrandoPRE2006,ref:GarciaMarchPHYSD2009}. It was shown that the values of $m$ are restricted by the order of symmetry~\cite{ref:Ferrando2005PRL}, and hence:
\begin{equation}
m=
\left\{ \begin{array}{l@{\quad\quad}l}
0,\pm 1,\pm 2, \dots,\frac{N}{2} &  \mathrm{even \;} N \\
\\
0,\pm 1,\pm 2, \dots,\frac{N-1}{2} & \mathrm{odd \;} N.
\end{array}\right. \end{equation}
Also, it was  shown that every vortex presents a phase singularity of charge $m$ in the origin, where the charge  of a singularity is $v_j= 1/2\pi \oint_{\Gamma_j}\nabla  \theta\cdot d\mathbf{l} $ where $\Gamma_j$ is a closed path that encircles only this singularity~\cite{ref:GarciaMarchPRA2009}. It was also shown that one can relate the angular momentum $\ell$ of a circularly symmetric vortex struck by a discretly symmetric potential of order $N$ with the angular pseudomomentum $m$ of the wave function in the discrete  symmetry media by the transformation rule~\cite{ref:Ferrando2005PRL,ref:PerezGarcia2007,ref:GarciaMarchPHYSD2009}:
\begin{equation}\label{transmutationrule} 
\ell-m=kN,
\end{equation}
 where $k$ is an integer.   The process of transformation has been described microscopically as the disintegration of the highly charged vortex into a number of smaller vortices~\cite{ref:Zacares2009}. In the axis of symmetry, a vortex of charge $m$ remains, according to~\cite{ref:GarciaMarchPRA2009}, while the integer $k$ in Eq.~\eqref{transmutationrule}  is related to the number of rings of single charged vortices emerging from the axis~\cite{ref:Zacares2009}. For example, consider a  circularly symmetric vortex with vorticity \(v=3\) being broken by an \(N=4\) symmetric impulse, the transformation rule in Eq.~\eqref{transmutationrule}, gives \(m=-1\) with $k=1$, and the central singularity has charge $-1$. In this case, the initial vortex will result in 5 singularities after symmetry breaking. One stays at the origin, with a new charge of \(v_j=-1\), and one ring of four  \(v_j=+1\) charged singularities  comes symmetrically off-axis, as represented in Fig.~\ref{fig:Fig1}. Here we are interested in the trajectories followed by these ejected daughter vortices after emerging from the axis. For illustration purposes, in the following we will consider only this particular case, even though all results can be extended to an arbitrary charge of the initial vortex and symmetry order of the impulse.

\section{Propagation of a vortex after a symmetry breaking impulse}
\label{sec:propagation}

In this section as well as Sec. \ref{sec:trajectories} we consider the linear case, $g=0$, and we use the substitutions
\begin{equation}
\omega t=\tau,\:L \equiv \sqrt{\frac{\hbar}{m\omega}},\:{\bf x}= {\bf \tilde{x}}/L,
\end{equation}
and we take \(t_0=0\). We consider the initial normalized vortex of charge $\ell=3$ given by 

\begin{align}\label{wave}\nonumber
\phi_{n,\;m}({\bf x})&=\sqrt{\frac{1}{6}}(x+iy)^3\frac{1}{\sqrt{2^{n+m}n!m!\pi }}\\
&\times \exp{\left[-\frac{x^2+y^2}{2 }\right]}H_n(x)H_m(y),
\end{align}
which is a solution of Eq.~\eqref{Schrodinger}. In the following we consider the vortex with charge $\ell=3$ with less energy, i.e.,  we set \(n=m=0\).  As shown in Appendix~\ref{app:evolutionoperator}, the amplitude of the vortex wavefunction after the action of the impulse will be given by
\begin{equation}\label{phibar}
\bar{\phi}(\chi)=e^{i\Delta V_0(\chi)\Delta \tau}\phi(\chi),
\end{equation}
where \(\phi(\chi)\) represents the vortex wavefunction before the impulse~\cite{ref:Andrews2008} and we use the complex notation \(\chi=x+iy\), and \(\Delta\tau\) is the duration of the impulse. 
Using the Taylor expansion of the evolution operator given in Eq.~\eqref{exptaylor} of Appendix~\ref{app:SBpotential} gives us 
\begin{equation}
\bar{\phi}(\chi)=e^{i \Delta \tau (v_0 \chi^4 +v_1 \chi^{*4})}\phi(\chi).
\end{equation}

If we carry out another Taylor expansion for the exponential, we get the expression for \(\bar{\phi}(\chi)\) that will be used for the propagation in the final medium
\begin{equation}
\bar{\phi}(\chi)=(1+iv_0\Delta\tau \chi^4 +iv_1\Delta\tau \chi^{*4})\phi(\chi).
\end{equation}
Inserting the initial state \(\phi_{nm}\) given in Eq.~\eqref{wave} into the previous equation, we arrive at the final form of our wavefunction after symmetry breaking.
\begin{align}\nonumber
\bar{\phi}({\bf x})=&\left[1+iv_0 (x+iy)^4 +iv_1 (x-iy)^{4}\right]\\\nonumber
&\times\sqrt{\frac{1}{6}}(x+iy)^3 \frac{1}{\sqrt{2^{n+m}n!m!\pi }}\\
&\times \exp{\left[-\frac{x^2+y^2}{2 }\right]}H_n(x)H_m(y),
\label{eq:initialcond}
\end{align}
where we have absorbed the length of the impulse, \(\Delta\tau\), into the constants \(v_0\) and \(v_1\) such that \(v_0 \Delta\tau= v_0\) and \(v_1 \Delta\tau= v_1\).
The two-dimensional harmonic oscillator propagator given by
\begin{align}\label{propagatorphi}\nonumber
&\psi({\bf x},\:t)=\frac{1}{2 i \pi \sin \tau}\exp{\left[\frac{i \cos\tau(x^2 +y^2)}{2 \sin\tau}\right]}
\iint dx_0dy_0\,\\
& \psi(x_0,\:\tau_0)
\exp{\left[\frac{i}{2 \sin\tau}\left[(x_0^2+y_0^2)\cos\tau-2(xx_0+yy_0)\right]\right]},
\end{align}
valid for \(t\ge 0\), is then used to evolve the initial function Eq.~\eqref{eq:initialcond} in the transverse plane. Taking as the initial state \(\phi_{00}\), we get for the vortex field after symmetry breaking
\begin{align}\label{phicomplex}
\phi_{\mathrm{c}}&(\chi,\;\tau)=e^{-8i\tau-|\chi|^2/2}\sqrt{\frac{\pi}{6}}\left\{e^{4i\tau}\chi^3 +iv_0 \chi^7\right.\\\nonumber
&+ iv_1\chi^*\left[-24+24e^{6i\tau}+|\chi|^2(|\chi|^2-6)^2\right.\\\nonumber
&\left.\left. +36e^{4i\tau}(|\chi|^2-2)+12e^{2i\tau}\left\{6+|\chi|^2(|\chi|^2-6)\right\}\right]\right\}.
\end{align}

\begin{figure}[h]
\begin{center}
\includegraphics[width=\columnwidth]{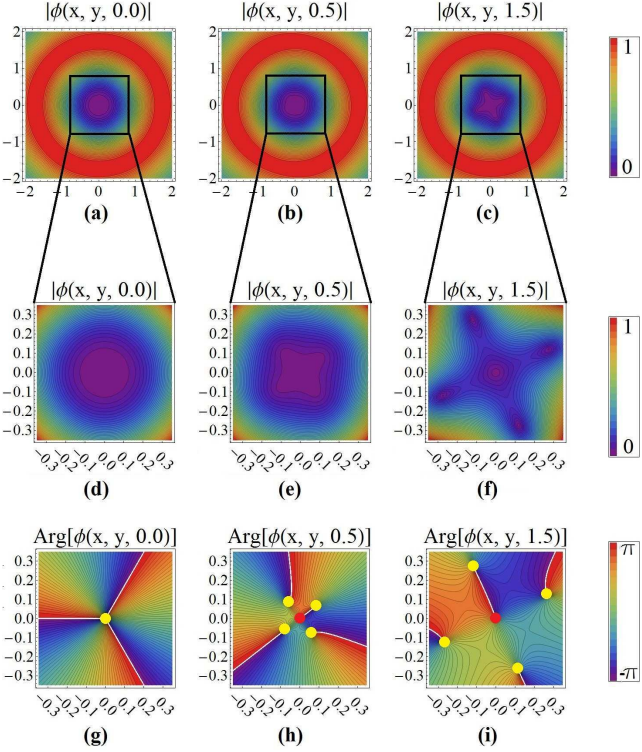}
\caption{  \label{fig:fig2}{The wave function $\phi(x,\:y,\:\tau)$ for different times}. (a) to ((f)  amplitude and a corresponding zoom for different time slices. (g) to (i) corresponding amplitude. Red dot represents position of negatively charged singularity at the origin, yellow dots represent the positions of the single positively charged singularities.}
\end{center}
\end{figure}

Let us write Eq.~(\ref{phicomplex}) as
\begin{align}\label{Aphi}
\phi_{\mathrm{c}}(\chi,\;\tau)=&e^{-8i\tau-\frac{|\chi|^2}{2}}\sqrt{\frac{\pi}{6}}\\ \nonumber
&\left(A_0(\tau) \chi^3 +A_+ \chi^7+A_-(|\chi|,\;\tau) \chi^*\right),
\end{align}
where
\begin{equation*}\label{Aplus}
A_+=iv_0,\,\,\,A_0(\tau)=e^{4i\tau},
\end{equation*}
and
\begin{align}\label{Aminus}
&A_-(|\chi|,\;\tau)=iv_1(-24+24e^{6i\tau}+|\chi|^2(|\chi|^2-6)^2\\\nonumber
& +36e^{4i\tau}(|\chi|^2-2)+12e^{2i\tau}(6+|\chi|^2(|\chi|^2-6))).
\end{align}

The expression in~\eqref{Aphi} has the form predicted by our previous symmetry arguments~\cite{ref:GarciaMarchPRA2009} since it can be written as
\begin{align}\label{Fphi}\nonumber
\phi(\chi,\;\tau)&=\sqrt{\frac{\pi}{6}}e^{-8i\tau-|\chi|^2/2}\chi^*\\\nonumber
&\quad\quad\quad\quad\quad\left[\frac{A_+\chi^8+A_0(\tau)\chi^4}{|\chi|^2}+A_-(|\chi|,\;\tau)\right]\\
&=\chi^* F(\chi,\;\tau),
\end{align}
where we have used the identities \(\chi^7/\chi^*=\chi^8/|\chi|^2\) and \(\chi^3/\chi^*=\chi^4/|\chi|^2\), and where
\begin{eqnarray}\label{F}
F(\chi,\;\tau)=&\sqrt{\frac{\pi}{6}}e^{-8i\tau-\frac{|\chi|^2}{2}}\\\nonumber
&\left[\frac{A_+\chi^8+A_0(\tau)\chi^4}{|\chi|^2}+A_-(|\chi|\;\tau)\right].
\end{eqnarray}

It becomes immediately apparent that \(F(\chi,\;\tau)\) is \(C_4\) invariant due to the dependence on only \(\chi^4\) and \(\chi^8\) terms. Because \(F(\chi,\;\tau)\) is \(C_4\) invariant, 
\begin{equation}
\phi(\epsilon \chi,\;\tau)=\epsilon^{-1}\phi(\chi,\;\tau),
\end{equation}
where \(\epsilon=e^{i\pi/2}\) is the elementary rotation of 4th order. Thus, as expected from the analysis in Ref.~\cite{ref:FerrandoPRE2006,ref:PerezGarcia2007,ref:GarciaMarchPHYSD2009,ref:GarciaMarchPRA2009}, using the transformation rule, the solution preserves the winding number \(m=-1\) for the center singularity. 

In Fig.~\ref{fig:fig2} (a), (b), and (c) we represent the amplitude of this function for different times.  A closer view of these amplitudes is shown in Fig.~\ref{fig:fig2} (d), (e), and (f), while Fig.~\ref{fig:fig2} (g), (h), and (i) show the phase for the same times and also near the origin. In this last figures the position of the singularities is highlighted. The positively charged singularities are ejected from the origin and follow some trajectory. In Sec. \ref{sec:trajectories}, we find the expressions for these trajectories. In Sec. \ref{sec:eom} we will determine the effective forces acting on these singularities.

\section{Trajectories of the phase singularities}
\label{sec:trajectories}

Let us obtain the trajectories followed by the phase singularities by finding the zeros of the complex wavefunction \(\phi(\chi,\;\tau)\). From \eqref{Fphi}, we see that there are two situations when \(\phi(\chi,\;\tau)=0\), when \(\chi^*=0\) and when \(F(\chi,\;\tau)=0\). 

For the former situation, we study the behavior of the wavefunction near the origin by developing \(\phi(\chi,\;\tau)\) in a Taylor series around \(\chi=0\), obtaining:
\begin{equation}
\phi(\chi,\;\tau)\approx 32e^{-5i\tau}\sqrt{6\pi}\sin^3(\tau) v_1 \chi^*.
\end{equation}
Evidently, the singularity at the origin is due to the symmetry breaking of the initial vortex into \(C_4\), as seen by the dependence on the symmetry breaking parameter \(v_1\). We see again that this singularity has winding number \(m=-1\), as evidenced by the factor of \(\chi^*\). If we set \(v_1=0\), then \(A_-(|\chi|,\;\tau)=0\), and the expansion about \(\chi=0\) is instead
\begin{equation}
\phi(\chi,\;\tau)\approx\sqrt{\frac{\pi}{6}}e^{-4i\tau}\chi^3,
\end{equation}
which preserves the initial winding number of \(l=3\), as seen by \(\chi^3\).

The latter type of phase singularity, when \(F(\chi,\;\tau)=0\), is more difficult to analyze because we have to work with the complex roots of the nonlinear equation \(F(\chi,\;\tau)=0\). This is the same as solving the equation
\begin{equation}\label{As}
A_+ \chi^8 +A_0(\tau)\chi^4+|\chi|^2 A_-(|\chi|,\;\tau)=0.
\end{equation}
To make the calculation easier, we assume that the two symmetry breaking parameters are equal. Thus, we take \(v_0= v_1=v\). 

If we go to the \(v=0\) limit, we see that \(A_+=0\) and \(A_-(|\chi|,\;\tau)=0\). For \(F(\chi,\;\tau)=0\) to be true in this limit, it follows that as \(v\rightarrow 0\), \(A_0(\tau) |\chi|^4\rightarrow 0\), and therefore \(\chi\rightarrow 0\), leading to the conclusion that \(\chi=\chi(v)\), and the statement that in the \(v\ll1\) regime, \(|\chi|\ll1\).

For small values of \(\chi\), the first terms that reappear in \eqref{As} are those in \(A_-(|\chi|,\;\tau)\) that depend on \(|\chi|^2\). Due to \(|\chi|\) being much less than 1, it follows that \(|\chi|^2>|\chi|^4>|\chi|^8\). By expanding out \(A_-(|\chi|,\;\tau)\), we see that the \(|\chi|^2\) term is
\begin{align}\nonumber
\lim_{v\ll 1}A_-(|\chi|,\;\tau)&\approx |\chi|^2(-24 i v+72i e^{2i\tau}v\\
&-72ie^{4i\tau}v+24ie^{6i\tau}v).
\end{align}
Using this approximation, we can instead solve the equation
\begin{equation}
F\approx A_0(\tau) \chi^4+ A_{-v\ll1}(|\chi|,\;\tau)=0,
\end{equation}
where we have kept only the nonzero terms from the \(v\ll 1\) limit.

\begin{figure}[t]
\begin{center}
\includegraphics[width=2.75in]{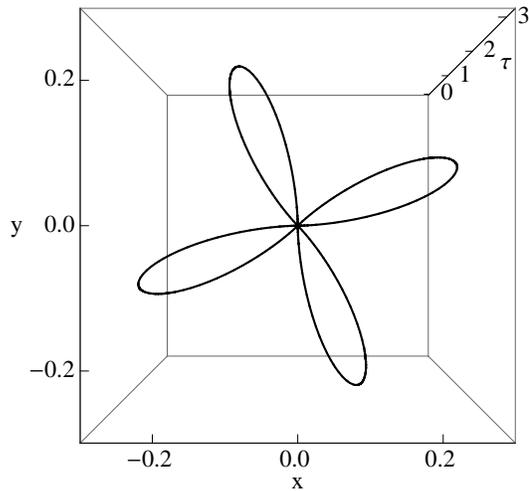}
\caption{\label{fig:fig3}{Trajectories followed by ejected daughter singularities after symmetry breaking. }}
\end{center}
\end{figure}

Thus, to order \(|\chi|^2\),
\begin{align}\nonumber
&e^{4i\tau}\chi^4+|\chi|^2(-24 i v+72i e^{2i\tau}v\\
&-72ie^{4i\tau} v+24ie^{6i\tau}v)=0.
\end{align}
If we solve for \(\chi^4\),
\begin{align}\label{p}\nonumber
\chi^4&=\frac{24 i v-72i e^{2i\tau}v+72ie^{4i\tau} v-24ie^{6i\tau}v}{e^{4i\tau}}|\chi|^2\\
&\equiv v p(\tau) |\chi|^2.
\end{align}
The simplest mathematical object to calculate now is \(|\chi|\). This is done by taking the modulus of the previous expression and dividing by \(|\chi|^2\). We obtain
\begin{equation}
|\chi|^2= 192  v_1 \sin^3(\tau).
\end{equation}
The equation above provides the evolution of the radial coordinate of the off-axis phase singularities after the symmetry-breaking impulse has been applied. Recall that \(|\chi(\tau)|^2=x(\tau)^2+y(\tau)^2=r(\tau)^2\) so that in polar coordinates the radius of the phase singularity trajectory is given by
\begin{equation}\label{r}
R(\tau)\approx 8\sqrt{3}(v_1 \sin^3(\tau))^{1/2}.
\end{equation}
To find \(\theta(\tau)\), we need to look back at \eqref{p}. If we rewrite \(\chi\) and \(p\) in modulus-argument complex form, \(\chi\) becomes \(|\chi|e^{i\theta}\) and \(p\) becomes \(|p|e^{i\gamma}\). Equation \eqref{p} becomes
\begin{equation}
|\chi|^2e^{i4\theta}=v |p|e^{i\gamma}.
\end{equation}
We saw in \eqref{p} that \(|\chi|^2=vp\), so the previous equation becomes
\begin{equation}
e^{i4\theta}=e^{i\gamma}= 4\theta=\gamma+2n\pi.
\end{equation}
Thus, the evolution of the polar coordinates of the phase singularities is provided by the phase of \(p(\tau)\). From \eqref{p},
\begin{equation}
p(\tau)=\frac{24 i-72i e^{2i\tau}+72ie^{4i\tau}-24ie^{6i\tau}}{e^{4i\tau}}.
\end{equation}
The phase of \(p(\tau)\) is found by taking the arctangent of \(p(\tau)\). This gives us
\begin{align}\label{theta}\nonumber
\theta(\tau)&=\frac{1}{4}\gamma=\frac{1}{4}\left[2n\pi+\arctan\left(\frac{ \sin^4(\tau)}{- \cos(\tau) \sin^3(\tau)}\right)\right]\\
&= \theta(\tau)\approx \frac{n\pi}{2}-\frac{\tau}{4}.
\end{align}
 The analytic trajectories calculated above are represented in  Fig.~\ref{fig:fig3}. 

\section{Equations of Motion}\label{sec:eom}
 In this section, we analyze the trajectories found previously to determine the effective forces acting on the singularities during and after symmetry-breaking. The equation of motion for the complex coordinate \(R(\tau)e^{i\theta(\tau)}\) shows that the system undergoes nontrivial motion corresponding to a harmonic oscillator with complex and time-dependent frequency. This leads us the discovery of an effective singular force that causes the daughter singularities to initially repel. Due to the complex nature of this effective potential, we show that the system of vortices is non-conservative. We then describe the torque about the origin acquired by the singularities during symmetry-breaking. This torque arises as the product of the Magnus force.
\subsection{Radial and Angular Equations of Motion}\label{sec:randaeom}
Now that we have expressions for \(R(\tau)\) and \(\theta(\tau)\), we can find the equations of motion for the off-axis singularities. By taking the derivatives of \(R(\tau)\), we see that the velocity and acceleration in the radial component can be expressed as
\begin{align}
\dot{R}(\tau)=&12\sqrt{3}\cos(\tau)(v_1 \sin(\tau))^{1/2}\\
\ddot{R}(\tau)=&3\sqrt{3}(3\cos(2\tau)-1)( v_1 \csc(\tau))^{1/2}.
\end{align}
By taking the derivatives of \(\theta(\tau)\), we see that the angular velocity is constant, $\dot{\theta}(\tau)=-\frac{1}{4}$, and therefore the angular acceleration is zero.

However, the fact that there is an angular velocity at all tells us that the singularities have acquired a torque about the axis at some point in their creation and propagation. If we recombine \(R(\tau)\) and \(\theta(\tau)\) into the complex coordinate \(\chi(\tau)=R(\tau)e^{i \theta(\tau)}\) once again, we can study the behavior of the singularities immediately after symmetry breaking.

If we Taylor expand equations \eqref{r} and \eqref{theta} around \(\tau=0\), i.e immediately after symmetry breaking, we get
\begin{equation}\label{rapprox}
R(\tau)\approx8\sqrt{3}\sqrt{v}\tau^{3/2},
\end{equation}
and
\begin{equation}
\theta(\tau)\approx\frac{\pi}{4}-\frac{\tau}{4}.
\end{equation}
The previous expansions give us the complex coordinate \(\chi(\tau)\) right after symmetry breaking, such that
\begin{equation}\label{chiapprox}
\chi(\tau)\approx8\sqrt{3 v}\;\tau^{3/2}e^{\frac{i}{4}(\pi-\tau)}.
\end{equation}
We now proceed to derive the equation of motion associated to \eqref{chiapprox}. If we differentiate \eqref{chiapprox} with respect to \(\tau\),  we see that
\begin{equation}
\chi'(\tau)=\left(\frac{3}{2\tau}-\frac{i}{4}\right)\chi(\tau).
\end{equation}
A second derivative of \eqref{chiapprox} will provide us with the equation of motion in complex notation:
\begin{equation}\label{chidoublespot}
\chi''(\tau)=\frac{12-\tau(12i+\tau)}{16\tau^2}\chi(\tau).
\end{equation}
If we let \(\Omega_0^2=\frac{1}{16}-\frac{3}{4\tau^2}\), and \(\Omega_1^2=\frac{3}{4\tau}\), we can rewrite \eqref{chidoublespot} as 
\begin{equation}
\chi''(\tau)=-(\Omega_0^2+i\Omega_1^2)\chi(\tau).
\end{equation}
Evidently, the phase singularities experience a nontrivial type of force. The previous equation represents a special type of harmonic oscillator in which the frequency is both complex and time dependent. Since the frequency is complex, we do not expect the system to be conservative. We can prove this statement by manipulating \eqref{chidoublespot} and its conjugate in the same manner we would do to establish conservation of energy in a standard harmonic oscillator. First, we write the conjugate of \eqref{chidoublespot}:
\begin{equation}\label{chidoublespotprime}
\chi''^*(\tau)=-(\Omega_0^2-i\Omega_1^2)\chi^*(\tau).
\end{equation} 
Next, we multiply \eqref{chidoublespot} by \(\chi'^*(\tau)\) and \eqref{chidoublespotprime} by \(\chi'(\tau)\) and add the two resulting equations to obtain
\begin{align}\nonumber
\chi''(\tau)&\chi'^*(\tau)+\chi'(\tau)\chi''^*(\tau)=\\ \nonumber
&=-(\Omega_0^2+i\Omega_1^2)\chi(\tau)\chi'^*(\tau)+\chi'(\tau)(-\Omega_0^2+i\Omega_1^2)\chi^*(\tau)\\\nonumber
&=-\Omega_0^2\chi(\tau)\chi'^*(\tau)-i\Omega_1^2\chi(\tau)\chi'^*(\tau)-\Omega_0^2\chi'(\tau)\chi^*(\tau)\\ \nonumber
&\;\;\;\;+i\Omega_1^2\chi'(\tau)\chi^*(\tau) \\\nonumber
&=-\Omega_0^2(\chi(\tau)\chi'^*(\tau)+\chi'(\tau)\chi^*(\tau))+i\Omega_1^2(\chi'(\tau)\chi^*(\tau)\\ \nonumber
&\;\;\;\;-\chi(\tau)\chi'^*(\tau)).
\end{align}

We immediately recognize that the left hand side and the first term of the right hand side are total derivatives. If we rewrite the total derivatives, we get
\begin{align}\label{derivatives}\nonumber
\frac{d}{d\tau}(\chi'(\tau)\chi'^*(\tau))&=-\Omega_0^2\;\frac{d}{d\tau}(\chi(\tau)\chi^*(\tau))\\ 
&+i\Omega_1^2(\chi'(\tau)\chi^*(\tau)-\chi(\tau)\chi'^*(\tau)).
\end{align}
Unfortunately, \(\Omega_0^2\) is time dependent, so we cannot just combine the total derivatives. Instead, we must subtract the term with \(\frac{d}{d\tau}\Omega_0^2\). The total derivative of the \(\Omega_0^2\) term is
\begin{equation}
\frac{d}{d\tau}(\Omega_0)^2 \chi(\tau)\chi^*(\tau))=\chi(\tau)\chi^*(\tau)\frac{d}{d\tau}\Omega_0^2+\Omega_0^2\frac{d}{d\tau}\chi(\tau)\chi^*(\tau).
\end{equation}
This allows us to rewrite \eqref{derivatives} as 
\begin{align}
&\frac{d}{d\tau}(\chi'(\tau)\chi'^*(\tau))+\Omega_0^2\frac{d}{d\tau}(\chi(\tau)\chi^*(\tau))\\ \nonumber
&=i\Omega_1^2(\chi'(\tau)\chi^*(\tau)-\chi(\tau)\chi'^*(\tau)),\\ \nonumber
\\ \nonumber
&\frac{d}{d\tau}\left(\chi'(\tau)\chi'^*(\tau)+\Omega_0^2\chi(\tau)\chi^*(\tau)\right)-\chi(\tau)\chi^*(\tau)\frac{d}{d\tau}\Omega_0^2\\ \nonumber
&=i\Omega_1^2(\chi'(\tau)\chi^*(\tau)-\chi(\tau)\chi'^*(\tau)).
\end{align}
If we replace \(\chi'(\tau)\) and \(\chi'^*(\tau)\) with their functional values, and evaluate the derivative of \(\Omega_0^2\), we obtain
\begin{align}
\frac{d}{d\tau}\left(\chi'(\tau)\chi'^*(\tau)+\Omega_0^2\chi(\tau)\chi^*(\tau)\right)&=\\
&\left(\frac{3}{2\tau^3}+\frac{\Omega_1^2}{2}\right)\chi(\tau)\chi^*(\tau).\nonumber
\end{align}

If we define the energy of the system the same way we would a typical harmonic oscillator, 
\begin{equation}
E=\frac{1}{2}\chi'(\tau)\chi'^*(\tau)+\frac{1}{2}\Omega_0^2\chi(\tau)\chi^*(\tau),
\end{equation}
it is clear that there is gain in the system. We can see the value of the gain by considering the derivative of the energy:
\begin{align}\label{energy}
\frac{dE}{d\tau}&=\frac{1}{2}\frac{d}{d\tau}\left(\chi'(\tau)\chi'^*(\tau)+\Omega_0^2\chi(\tau)\chi^*(\tau)\right)\\
&=\frac{1}{2}\left(\frac{3}{2\tau^3}+\frac{\Omega_1^2}{2}\right)\chi(\tau)\chi^*(\tau)\\
&=\frac{3}{4\tau}\left(\frac{1}{\tau^2}+\frac{1}{4}\right)|\chi(\tau)|^2\ge 0.
\end{align}
Thus, energy is not conserved by our equations of motion governing singularity or vortex motion.  However, the GPE does conserve energy.  Thus energy is being exchanged between the singularities and the remainder of the Bose-Einstein condensate described by the full GPE.

The presence of this effective harmonic motion explains why the post-symmetry breaking singularities expel from the origin. However, we still need to explain the effective torque that the singularities seem to experience. To understand this torque better, let us rewrite our complex coordinate in Cartesian coordinates via the definition of \(\chi(\tau)=x(\tau)+iy(\tau)\). This gives us
\begin{align}
\chi''(\tau)&=x''(\tau)+iy''(\tau)\\\nonumber
&=-(\Omega_0^2+i\Omega_1^2)(x(\tau)_iy(\tau))\\\nonumber
&=-(\Omega_0^2+i\Omega_1^2)x(\tau)-i(\Omega_0^2+i\Omega_1^2)y(\tau)\\\nonumber
&=-\Omega_0^2x(\tau)-i\Omega_1^2 x(\tau)-\Omega+0^2y(\tau)+\Omega_1^2y(\tau).
\end{align}
If we collect the real and imaginary parts, we arrive at
\begin{align}
x''(\tau)&=-\Omega_0^2x(\tau)+\Omega_1^2y(\tau)\\
y''(\tau)&=-\Omega_0^2 y(\tau)-\Omega_1^2 x(\tau).
\end{align}
We can write the previous equations in vector form as
\begin{equation}
{\bf r}\;''(\tau)=-\Omega_0^2{\bf r}(\tau)+\Omega_1^2 \begin{bmatrix}0&1\\-1&0\end{bmatrix}{\bf r}(\tau).
\end{equation}
In order to see how a torque comes into our system, we need to rewrite the \(\Omega_1^2\) matrix term in three-dimensions (3D). To do this, we construct the external 3D vector \(\Lambda=(0,\:0,\:\Omega_1^2)\) such that
\begin{equation}
{\bf r}\times\Lambda=\begin{vmatrix} {\bf i}&{\bf j}&{\bf k}\\x&y&z\\0&0&\Omega_1^2\end{vmatrix}=\Omega_1^2(y,\:-x,\:0)=\Omega_1^2\begin{bmatrix}0&1\\-1&0\end{bmatrix}{\bf r}_T(\tau),
\end{equation}
where \({\bf r}_T(\tau)\) is the transverse plane and {\bf i, j, k} are unit vectors, which we are working in. Therefore, the equation of motion for phase singularities can be represented in 3D, although the motion is restricted to a two-dimensional plane \({\bf r}(\tau)=(x,\:y,\:0)\)\footnote{Many authors describe the evolution of phase singularities directly through these loops. }. We write our 3D representation as
\begin{equation}\label{rdoublespot}
{\bf r}\;''(\tau)=-\Omega_0^2{\bf r}(\tau)+{\bf r}(\tau)\times\Lambda
\end{equation}
This equation of motion shows the simultaneous presence of a harmonic force and an external force associated with a torque. The fact that the latter is associated with a torque can be checked by calculating its effect on the angular momentum of the phase singularity -- \({\bf L}={\bf r}\times {\bf r'}\). If we look at the derivative of the angular momentum, we see that
\begin{equation}
\frac{d{\bf L}}{d\tau}=\frac{d}{d\tau}({\bf r}\times {\bf r'})={\bf r'}\times{\bf r'}+{\bf r}\times{\bf r''}={\bf r}\times{\bf r''}.
\end{equation}
If we evaluate this cross product using our expression for \({\bf r''}(\tau)\) in \eqref{rdoublespot}, 
\begin{align}\nonumber
{\bf r}\times{\bf r''}&={\bf r}\times(-\Omega_0^2{\bf r}+({\bf r}\times\Lambda)\\\nonumber
&=-\Omega_0^2({\bf r}\times{\bf r})+{\bf r}\times({\bf r}\times\Lambda)\\
&={\bf r}\times({\bf r}\times\Lambda).
\end{align}
Using the vector triple product \cite{ref:Kiyosi1993}, we obtain
\begin{equation}
{\bf r}\times{\bf r''}={\bf r}({\bf r}\cdot\Lambda)-\Lambda({\bf r}\cdot{\bf r}).
\end{equation}
Because \(\Lambda\) is defined only to have a \(z\) component, and our position vector is two-dimensional, the dot product of \({\bf r}\) with \(\Lambda\) vanishes, leaving
\begin{align}\nonumber
{\bf r}\times{\bf r''}&=-\Lambda({\bf r}\cdot{\bf r})\\\nonumber
&=-\Lambda|{\bf r}|^2\\
&=(0,\;0\;,-\Omega_1^2|{\bf r}_T|^2).
\end{align}

Finally, we arrive at 
\begin{equation}\label{torque}
\frac{d{\bf L}}{d\tau}=\tau=\left(0,\;0,\;-\Omega_1^2|{\bf r}_T|^2\right)=\left(0,\;0,\;-|{\bf r}_T|^2\frac{3}{4\tau}\right).
\end{equation}
The previous equation shows that the angular momentum has variance only in the \(z\) direction, which means there is a torque that causes rotation in the \(x,\;y\) plane, as we expect. Because the value of the torque is negative, our singularities rotate about the origin in a clockwise manner, as our trajectories in Sec.~\ref{sec:trajectories} were seen to do in Fig. \ref{fig:fig2}. In Appendix~\ref{sec:energy}, we check that the calculation of the energy is correct using our 3D formalism. 

Let us note that all of these results apply, when properly rotated, to any of the four phase singularities moving away from the center of symmetry. This is due to the four-fold symmetry of our solutions and it is reflected in the four solutions that we have for the angular coordinate \(\theta(\tau)\) in \eqref{theta}.

\subsection{Dynamics of Phase Singularities in Free Space after Rotational Symmetry Breaking}

In Sec. \ref{sec:randaeom} we derived the equation of motion for the four phase singularities that arise immediately after symmetry breaking by a discretely symmetric impulse. We found that the breaking of rotational symmetry causes a  vortex to cluster in a central singularity carrying topological charge equal to the angular pseudo-momentum \(m\) and a ``wave" of \(N\) (\(N\) being the order of symmetry of the impulse) single charged phase singularities with particle-like motion moving away from the symmetry axis. The dynamics of these phase singularities as point-like particles is described by the equation of motion in \eqref{rdoublespot} (for the case \(N=4\)). This equation is very interesting because it shows that, despite the wave function describing the propagation of matter corresponds to linear harmonic propagation, the clustered phase singularities do not move as harmonic oscillators. In fact, right after the action of the impulse, they experience two types of forces, as described by the right-hand side of \eqref{rdoublespot}:
\begin{itemize}
\item A harmonic repulsive force given by \(\Omega_0^2{\bf r}(\tau)\).
\item A rotational force \({\bf F}=({\bf r}(\tau)\times\Lambda)\) generating a torque \({\bf M}=-|{\bf r}|^2\Lambda\).
\end{itemize}
Both forces have a peculiar behavior. Let us analyze them separately.

\subsubsection{Effective Harmonic Potential}
This effective potential is crucial because it is responsible for the dissociation of the initial highly-charged vortex with topological charge \(\ell=3\) into the central singularity of charge \(m=-1\) and four vortices of charge \(v_j=+1\). If the interaction was attractive, the four vortices would remain at the origin (the center of symmetry) since both the initial position and initial velocity are zero.  However, we find that the interaction is repulsive because \(\Omega_0^2=\frac{1}{16}-\frac{3}{4\tau^2}<0\) for small values of \(t\). Nevertheless, a repulsive harmonic interaction is not enough to guarantee the motion of the broken singularities away from the origin since their position and velocity are initially zero. They would remain there in a situation of unstable equilibrium since the force upon them would be zero. Something else is needed to trigger the expansive motion of the broken singularities. The mechanism is the existence of a nonzero, in this case singular, repulsive potential at \(t=0\). 
\begin{equation}
|{\bf F}_H|=\left\lvert\left(\frac{1}{16}-\frac{3}{4\tau^2}\right)\right\rvert r(\tau)\sim \frac{1}{\tau^2}\tau^{3/2}=\frac{1}{\sqrt{\tau}}\xrightarrow{t\to 0}\infty.
\end{equation}
If we analyze the form of the effective harmonic potential for small values of \(t\), we see from \eqref{energy} that
\begin{equation}
V_H({\bf r})=\frac{1}{2}\Omega_0^2|{\bf r}|^2\approx-\frac{3}{4\tau^2}|{\bf r}|^2\:\:t\ll1,
\end{equation}
indicating the presence of a singular repulsive potential at \(t=0\). The curvature of the quadratic potential is, thus, infinite and negative right after the symmetry is broken, so the force on the escaping singularities is non-zero when they are located at the origin when \(t=0\). This singular potential is the reason why the singularities start to move away from the center of symmetry. The fact that the potential and force are singular at \(t=0\) does not produce any issues in the velocity and position of the fleeing singularities when \(t=0\) because the acceleration, which has the form \(r''(\tau)\sim 1/\sqrt{\tau}\) has first and second integrals of the form:
\begin{equation}
r'(\tau)\sim\sqrt{\tau}+C\textrm{          and          }r(\tau)\sim \tau^{3/2}+C',
\end{equation}
which are both finite at \(t=0\) and compatible with the initial condition \(r'(0)=0\) and \(r(0)=0\) when the constants are taken to be zero. 

\subsubsection{Torque}

As seen in \eqref{torque}, there is an \(r\) dependence in the torque that the singularities experience around the origin once symmetry is broken. The torque is zero when \(t=0\) since \(\tau\sim r^2\xrightarrow{t\to0}0\) due to the initial condition of \(r(0)=0\). Thus, the singularities must start moving away from each other, making \(r\ne 0\), before the external torque can take effect. This allows us to conclude that the singular repulsive effective harmonic potential acts on the singularities before they can acquire any angular momentum.

As our vortices acquire a linear velocity away from the origin, they become subject to the Magnus effect. This effect creates a force perpendicular to the direction of motion according to \(\vec{F}=S(\vec{\omega}\times\vec{v})\), where \(S\) is a property of the medium the vortex is traveling through, \(\vec{\omega}\) is the angular rotational velocity of the spinning object, and \(\vec{v}\) is the linear velocity. This perpendicular Magnus force causes the vortices to follow a curved path. If we evaluate the expression for the Magnus force, we see that
\begin{align}\nonumber
\vec{F}&=S(\vec{\omega}\times\vec{v})\\\nonumber
&=S\left( 0,\;0,\;\omega\right)\times\left( \frac{3}{4\tau}x,\;\frac{3}{4\tau}y,\;0\right)\\
&=S\left( -\frac{3\omega}{4\tau}y,\frac{3\omega}{4\tau}x,0\right),
\end{align}
where the velocity vector was formed by taking the derivative of the position at small \(t\) given in \eqref{rapprox}. If we evaluate the torque associated with the Magnus force, we see that
\begin{align}\nonumber
\tau&=\vec{r}\times\vec{F}\\\nonumber
&=\left( x,\;y,\;0\right) \times S\left( -\frac{3\omega}{4\tau}y,\frac{3\omega}{4\tau}x,0\right)\\
&=\left( 0,\;0,\;\frac{3S\omega}{4\tau}|\vec{r}|^2\right),
\end{align}
which is consistent with our expression for the torque found in the previous section with \(S=-1\) and \(\omega=1\), verifying that the torque associated with the singularities after symmetry breaking is generated by the Magnus force.

Therefore, the dynamics of our singularities after symmetry breaking can be described as follows: First, the action of the symmetric impulse introduces an effective singular repulsive harmonic potential that splits \(N=4\) single phase singularities out of the original highly charged vortex. As these singularities begin to travel away from the origin, they gain angular momentum from the effective external torque and rotate around the axis of symmetry. Eventually, the effective harmonic potential is overpowered by the trapping potential, so the singularities travel back toward the origin, and settle into oscillatory motion about the origin. Eventually, they fuse back to the origin and reconstruct the initial vortex for the non-interacting case only. 

We can also find an expression for the maximum radius the singularities achieve as a function of \(v\), given by 
\begin{equation}
R_{\mathrm{max}}=8\sqrt{3v},
\end{equation}
confirming that the strength of the impulse directly affects the motion of the singularities.

\section{Trajectories of singularities in the presence of interactions}
\label{convergence}

We can compare the calculated trajectories from section \ref{sec:trajectories} with the actual minima of the wavefunction amplitude as well as numerical data generated by numerically solving the GPE~\eqref{Gross}. To begin, we compare the analytic solutions with the actual minima of the wavefunction to determine the accuracy of the calculated trajectories for various impulse strengths. Later, we compare the analytic trajectories, solved for a non-interacting BEC, and compare them to the numerical data for the same symmetry-breaking process in a weakly interacting BEC with various particle interaction strengths to determine the validity of our results in the nonlinear case. 

\subsection{Linear Comparison}
\begin{figure}[h]
\begin{center}
\includegraphics[width=\linewidth]{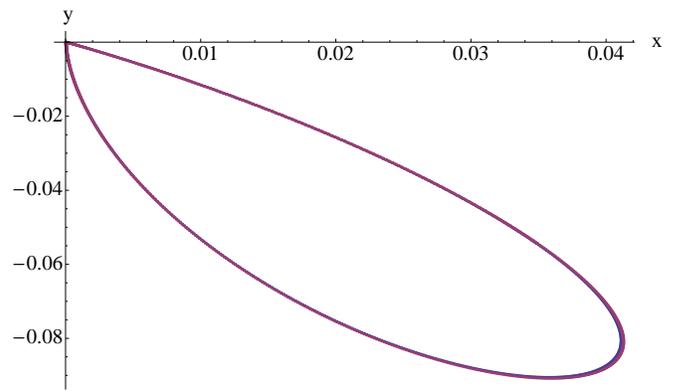}
\caption{  \label{fig:fig4}  (Color online). \emph{The calculated trajectories (pink) are plotted against the amplitude minima (blue) for discrete time steps, using a value of $v=0.00005$ for the area of the symmetry-breaking impulse. The approximated trajectories are a good description of the singularity motion for small impulse areas.}}
\end{center}
\end{figure}

To compare the analytic trajectories to their actual locations in the wavefunction, we must find a way to track the singularities. Due to the non-analyticity of the wavefunction, we must use the {\it Minimize} command in Mathematica in order to track the singularities for various time steps. Using a {\it Do} loop, we can append the location of the minima in the fourth quadrant to a list and plot the trajectories. As we do so, we can compare the calculated trajectories (in pink) with the located minima (blue) for \(v=0.00005\) in Fig. \ref{fig:fig4}. The value of \(v\) is a numerical representation of the impulse area, \(\Delta V_0\Delta t\), as described by the potential profile in \ref{system}.

\begin{figure}[h]
\begin{center}
\includegraphics[width=\linewidth]{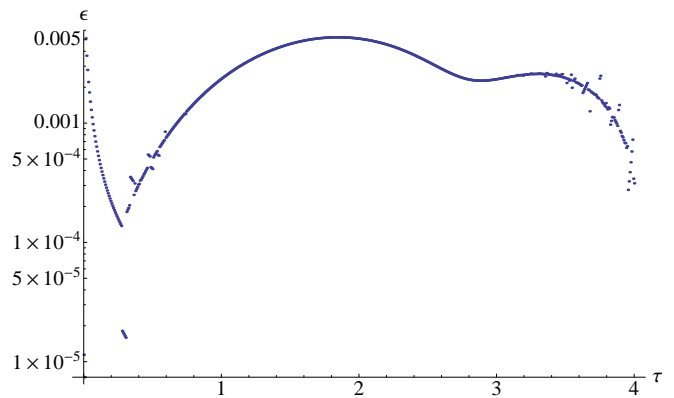}
\caption{  \label{fig:fig5}  (Color online). \emph{We plot the error on a logarithmic scale between the calculated trajectories and the local minima of the wavefunction amplitude for $v=0.00005$. The error is always less than 0.5\%. The scattered points at the beginning and end of the plot are due to the \textit{minimize} command in Mathematica being unable to distinguish between the four external singularities when they are extremely close to the origin. }}
\end{center}
\end{figure}

We can calculate the error between the two trajectories by using the formula
\begin{equation}
\varepsilon=\text{Log}_{10}\left|\frac{r_{\mathrm{calc}}-r_{\mathrm{theory}}}{\frac{1}{2}(r_{\mathrm{calc}}+r_{\mathrm{theory}})}\right|
\end{equation}
where \(r=x^2+y^2\). As one can see in Fig. \ref{fig:fig5}, the error is largest at the apex of the petal loop, but is still within \(0.5\%\) of the amplitude minima for \(v=0.00005\). The first few points in Fig. \ref{fig:fig5} have very large error due to the close proximity of the singularities immediately after symmetry breaking. The {\it minimize} command searches for a local minima, so when all four external singularities are very close to each other, the local minima could be from any of the singularities, increasing the error.

The error is insignificant until the outer edge of the petals. The increase in error is most likely due to the various approximation techniques used to calculate the analytical trajectories, one of which was working close to the origin. 

\begin{figure}[h]
\begin{center}
\includegraphics[width=\linewidth]{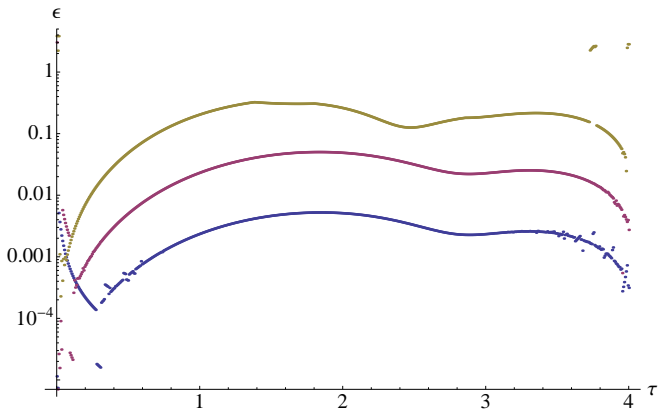}
\caption{  \label{fig:fig6}  (Color online). \emph{We plot the error for impulse areas of $v=0.005$ (yellow), $v=0.0005$ (pink), and $v=0.00005$ (blue). The error increases significantly as the duration of the impulse increases, but remains the lowest near the origin.}}
\end{center}
\end{figure}

We can increase the value of the impulse area, \(v\), to observe the loss of validity as the duration of the impulse increases. As one can see in Fig. \ref{fig:fig6}, the error significantly increases as the impulse area, \(v\) increases. The error becomes greatest near the apex of the petal structure, while still being within 5\% near the origin. The error rises above 5\% for times greater than \(\tau\approx0.5\) for the large impulse area, \(v=0.005\).

These results show that the analytic trajectories are the best approximations for very small impulse areas, but are still valid near the origin for larger impulse areas.

\subsection{Impulse Approximation}
\begin{figure}[h]
\begin{center}
\includegraphics[width=\linewidth]{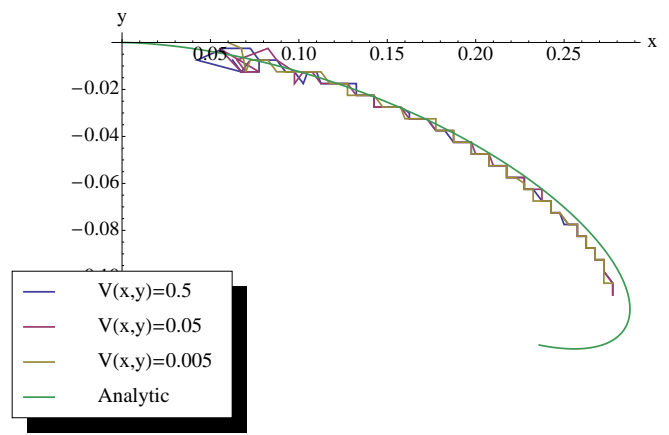}
\caption{  \label{fig:impulses}  (Color online). \emph{We plot the numerical trajectories for impulse durations of $\Delta t=0.1$, $\Delta t=0.01$, and $\Delta t=0.001$, and compare to analytic trajectories. We see no significant change as the duration is increased.}}
\end{center}
\end{figure}

To determine whether it is the duration of the impulse or the height of the impulse that affects the error, we include numerical integration studies similar to those in Sec: \ref{sec:numerics} but for \(g=0\). Three studies were evaluated for impulse area \(v=0.0005\), each with a different duration and height.
\begin{eqnarray}
V(x,\;y)&=&0.005\quad\Delta t=0.1\quad v=0.0005\\
V(x,\;y)&=&0.05\quad\Delta t=0.01\quad v=0.0005\\
V(x,\;y)&=&0.5\quad\Delta t=0.001\quad v=0.0005
\end{eqnarray}
If we plot each case, we see that there is no significant difference as the duration of the impulse is increased, as shown in Fig. \ref{fig:impulses}.

\begin{figure}[h]
\begin{center}
\includegraphics[width=\linewidth]{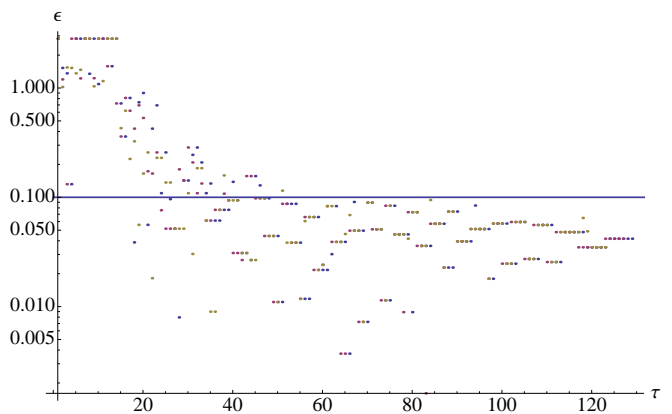}
\caption{  \label{fig:impulseerror}  (Color online). \emph{We plot the error between numerical trajectories for impulse durations of $\Delta t=0.1$, $\Delta t=0.01$, and $\Delta t=0.001$, and  analytic trajectories. We see no significant change as the duration is increased.}}
\end{center}
\end{figure}

If we calculate the error between the analytic and numerical trajectories, we see that the error stays below \(10\%\) once the singularities leave the origin. The large error near the origin is due to the closeness of all off-axis singularities to the central one, together with the impossibility of locating the singularities with an accuracy smaller than the grid spacing used in the numerical simulations of the Eq.~\eqref{Gross}. We plot the error on a logarithmic scale in Fig. \ref{fig:impulseerror}.

These results allow us to conclude that the actual duration of the impulse does not significantly change the dynamics of the system so long as the total impulse area is small. For the analytic trajectories, this means the approximation is valid to within \(5\%\) for impulse areas less than \(v\Delta \tau=0.0005\). For small areas, the impulse only serves to break the symmetry of the singularities, and becomes negligible if the duration is increased. In essence, a shallow potential for a longer time which is not governed by the impulse approximation has the same effect as an extremely strong potential for an infinitesimal amount of time, as long as the area \(v\Delta \tau\) remains small.

\subsection{Numerical Comparison - Nonlinear Case}\label{sec:numerics}
The GPE, given by \eqref{Gross},
was solved numerically for the same impulse used in the analytical analysis for various values of the nonlinearity, {\it g}. The nonlinearity depends explicitly on the scattering length between particles. The time at which the nonlinearity becomes significant is approximately 
\begin{equation} 
\tau_{\mathrm{nonlin}} = \omega t_{\mathrm{nonlin}} = \omega \frac{\hbar L}{N g}\, , 
\end{equation}
derived by units considerations from the renormalized effective 2D interaction strength\cite{carr2005c} $g=g_{\mathrm{2D}}\equiv \sqrt{8\pi\hbar^3\omega_z}{M} a_s \propto g_{\mathrm{3D}} $, with $\omega_z$ the transverse harmonic oscillator frequency, and $L_z\equiv \sqrt{\hbar/m\omega_z}$; we use simply ``$g$'' for our 2D effective interaction strength for simplicity of notation. We observe the numerical data for attractive nonlinearity and repulsive nonlinearity to see the structure of the vortex trajectories in each case. 
The effect we pursue is in the very core of the vortex, but there is an unavoidable limitation related to the grid spacing necessary to compute the minima of the wavefunction. Thus, the numerics have large error near the origin where the vortex cores initially overlap and again approach closely at later times.

\subsubsection{Repulsive Nonlinearity}
\begin{figure}[h]
\begin{center}
\includegraphics[trim =10mm 0mm 10mm 0mm, clip, width=\columnwidth]{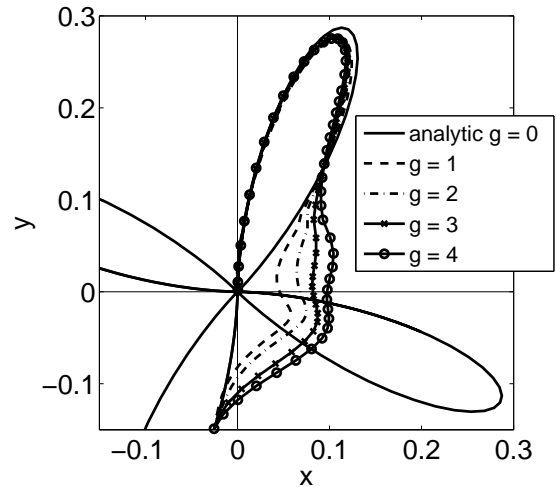}
\caption{  \label{fig:fig7}  (Color online). \emph{Comparison between repulsive nonlinear numerical data for $g=1,\;2,\;3,\;4$. As the nonlinearity increases, the singularities are less likely to return to the origin. The repulsive nonlinearity sets the trajectories in a completely different orbit than the nonlinear case of $g=0$. This repulsive motion is similar to a system of like-charged particles in that the singularities interact with each other before returning to the oscillatory path about the origin.}}
\end{center}
\end{figure}

Repulsive nonlinearity arises when the particles in a Bose-Einstein condensate interact with one another via a positive s-wave scattering length, corresponding to positive values of \(g\). As the nonlinearity becomes larger, the trajectories begin to interact at further distances from each other. This prevents the singularities from recombining at the origin, and instead sends the trajectories into repulsive motion, similar to the behavior of like-charged particles, before returning to the oscillatory path about the origin. Trajectories derived from numerical integration for various repulsive nonlinearities, \(g\), can be seen in Fig. \ref{fig:fig7}. 

By increasing the nonlinearity from \(g=0\) to \(g=1,\,2,\,3,\,4\), we see that when the singularities come back to the origin, the nonlinearity begins to show its effects, as seen by the paths taken by the numerical data. As seen in the previous figure, once nonlinearity is introduced, the singularities interact before traveling straight across the origin. In the non-interacting case, the singularities do not come back to the origin, but instead switch directly to another of the four loops. To show this, we plot in Fig. \ref{fig:fig7} the trajectories for all four singularities in linear case, while only the trajectory of one of them in the nonlinear ones. 

\subsubsection{Attractive Nonlinearity}
\begin{figure}[h]
\begin{center}
\includegraphics[trim =10mm 0mm 10mm 0mm, clip, width=\columnwidth]{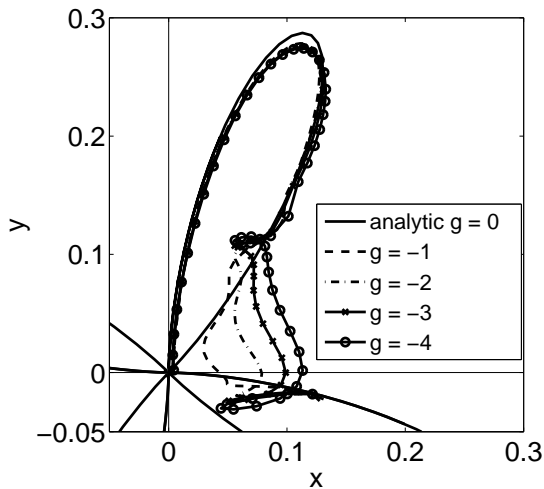}
\caption{  \label{fig:fig8}  (Color online). \emph{Comparison between attractive nonlinear numerical data for $g=-1,\;-2,\;-3,\;-4$. As the nonlinearity increases, the singularities are less likely to return to the origin. The negative nonlinearity sets the trajectories into orbital motion in the opposite direction as the nonlinear case, essentially "sling-shotting" around each other, causing the singularities to interact with each other before returning to the oscillatory path about the origin.}}
\end{center}
\end{figure}

Attractive nonlinearity corresponds to a negative value for the s-wave scattering length, resulting in negative values of the nonlinearity, \(g\). As the attractive nonlinearity becomes increasingly negative, the trajectories begin to interact at further distances from each other. This prevents the singularities from recombining at the origin, behaving similarly to a system of planets, essentially "sling-shotting" around each other before returning to their oscillatory paths. Trajectories calculated numerically for attractive nonlinearity can be seen in Fig. \ref{fig:fig8}.

By increasing the attractive nonlinearity from \(g=0\) to \(g=-1,\,-2,\,-3,\,-4\), we see that the nonlinearity begins to show its effects near the origin, as seen by the paths taken by the numerical data. The negative nonlinearity sets the trajectories into orbital motion, essentially "sling-shotting" around each other before returning to oscillatory motion. However, with the attractive nonlinearity, the trajectories return to the opposite orbit as with the linear \(g=0\) data. Again, in Fig. \ref{fig:fig8} we show the trajectories for all four singularities in the linear case, and only one of them in the nonlinear ones.

\section{Conclusions}
\label{sec:conclusion}
We have analytically described the equations of motion for the off-axis singularities that arise after the action of a symmetry-breaking impulse on an initial single highly charged vortex. For an initial vortex of vorticity \(\ell=3\) at the origin and a \(C_4\) discretely symmetric impulse, the symmetry of the initial vortex is broken into \(C_4\) as well. Four vortices with charge \(v_j=+1\) oscillate about the origin in a flowering pattern. A single vortex of charge \(v_j=-1\) remains stationary at the origin. 

All future evolution of the singularities is determined by the order of symmetry of the impulse. The singularities are imprinted by the impulse and ``remember" the effect of symmetry-breaking once back to an ordinary confining harmonic potential. It is interesting to note that the actual form of the impulse does not change the motion of the singularities. It is the order of symmetry that determines all future propagation patterns. 

The calculated trajectories of the off-axis singularities give rise to a blossoming structure. The singularities periodically oscillate about the origin, while rotating about the axis of symmetry. The disassociation of the initial highly-charged vortex into several smaller vortices is due to an effective singular repulsive harmonic potential that is introduced by the symmetry-breaking impulse. The singularities also acquire angular momentum around the axis of symmetry due to an external effective torque caused by the Magnus force. Once the effective repulsive potential is overpowered by the trapping harmonic potential, the singularities settle into an oscillatory pattern as expected in a harmonic trap.

The analytic trajectories were compared with the local minima of the wavefunction for impulse strengths of \(v=0.005\), \(v=0.0005\), and \(v=0.00005\). Comparison with the local minima showed the trajectories to be within \(0.5\%\) error for \(v=0.00005\), \(5\%\) error for \(v=0.0005\), and \(50\%\) error for \(v=0.005\). The increase in error as the impulse duration is increased is due primarily to approximations made in the analytical analysis. By superimposing the analytic trajectories with the local minima, we see that they are in agreement for small impulse strengths. The actual duration of the impulse does not significantly change the dynamics of the system so long as the total impulse area is small, less than \(V(x,\;y)\Delta t=0.0005\).

The initial break-up of the singularity is completely controlled by linear effects. It is only long-time behavior that requires full nonlinear analysis due to the interaction between particles in an interacting BEC.

We acknowledge support from the U.S. National Science Foundation (KAC and LDC), the Alexander von Humboldt foundation (LDC), and the Heidelberg Center for Quantum Dynamics (LDC). A.F. acknowledges  support by Contract No. TEC2010-15327. MAGM acknowledges support from Spanish ministry of Science and Education (MEC) and US Fulbright Commission.

\appendix

\section{Evolution operator}
\label{app:evolutionoperator}

For the potential profile given in~\eqref{profile}, we can decompose the evolution operator into three separate operators, one for each region, according to
\begin{equation}
e^{i\hat{H}\tau}=e^{i\hat{H}(\tau-\tau_1)}e^{i\hat{H}\Delta \tau}e^{i\hat{H}\tau_0}.
\end{equation}
Now, \(\hat{H}=\hat{H}_{HO}+V_{SB}\), where \(\hat{H}_{HO}\) is the initial Hamiltonian of the harmonic oscillator, and \(V_{SB}\) is the symmetry-breaking potential, both in dimensionless units. We define \(\hat{H}_0=\hat{H}_{HO}+V_0\) and \(\hat{H}_1=\hat{H}_{HO}+V_1\). According to the potential profile given in Eq.~\eqref{profile}, the evolution operator can be rewritten as 
\begin{equation}
e^{i\hat{H}\tau}=e^{i\hat{H}_1 (\tau-\tau_1)}e^{i(\hat{H}_0 +\Delta V_0)\Delta\tau}e^{i\hat{H}_0 \tau_0}.
\end{equation}
Let us analyze the evolution operator for the impulse. Since \(\Delta\tau\ll 1\), we can apply the Hausdorff-Campbell decomposition \cite{ref:Magnus1954} via the Zassenhaus formula to lowest order to get
\begin{align}\nonumber
e^{i(\hat{H}_0+\Delta V_0)\Delta \tau}&=e^{i\hat{H}_0 \Delta \tau}e^{i\Delta V_0 \Delta\tau}+O(\Delta\tau^2)\\
&=e^{i\Delta V_0 \Delta\tau}e^{i\hat{H}_0 \Delta\tau}+O(\Delta\tau^2),
\end{align}
where the two orders of the operators are possible since they commute with \(O(\Delta\tau^2)\). The full evolution operator is then given by
\begin{align}\nonumber
e^{i\hat{H}\tau}&=e^{i\hat{H}_1(\tau-\tau_1)}e^{i\Delta V_0 \Delta\tau}e^{i\hat{H}_0 \Delta\tau}e^{i\hat{H}_0 \tau_0}\\
&=e^{i\hat{H}_1(\tau-\tau_1)}e^{i\Delta V_0 \Delta\tau}e^{i\hat{H}_0(\tau_0+\Delta\tau)}.
\end{align}
If we take into account that \(\tau_1=\tau_0+\Delta\tau\), we finally write
\begin{equation}
e^{i\hat{H} \tau}=e^{i\hat{H}_1(\tau-\tau_1)}e^{i\Delta V_0 \Delta\tau}e^{i\hat{H}_0 \tau_1}.
\end{equation}
\\
If we apply this operator to an initial wave function, we see that
\begin{align}\nonumber
|\phi(\tau)\rangle&=e^{i\hat{H}_1(\tau-\tau_1)}e^{i\Delta V_0 \Delta\tau}e^{i\hat{H}_0 \tau_1}|\phi(0)\rangle\\
&=e^{i\hat{H}_1(\tau-\tau_1)}e^{i\Delta V_0 \Delta\tau}|\phi(\tau_1)\rangle.
\end{align}
It turns out that the presence of an impulse at time \(\tau_1\) only produces a multiplication by the diagonal operator in position space, \(e^{i\Delta V_0({\bf x}) \Delta\tau}\). If we define 
\begin{equation}\label{phibarpre}
\bar{\phi}(\tau_1)=e^{i\Delta V_0 \Delta\tau}|\phi (\tau_1)\rangle,
\end{equation}
the resulting amplitude can be propagated to future times using the harmonic oscillator Feynmann propagator in the final medium. 

\section{Symmetry breaking potential}
\label{app:SBpotential}

Close to the origin, \(|\chi|^2=x^2 +y^2 \rightarrow 0\) so we can perform a Taylor expansion of the evolution operator in \eqref{phibarpre} in the complex variable \(\chi\) and keep the lower order terms. Because of the \(C_N\) invariance of the potential, there are only two types of \(C_N\)-invariant products of \(\chi\) and \(\chi^*\) that can appear in this Taylor expansion: \(\chi\chi^*=|\chi|^2=x^2 +y^2,\:\chi^N,\) and \(\chi^{*N}\).

Here, we consider discrete rotational symmetry of order \(N=4\). If we perform a Taylor expansion on the arbitrary impulse function \(V(\chi)\) in both variables and keep the allowed terms mentioned previously, the potential of the impulse can be expanded to read
\begin{equation}
\Delta V_0(\chi)=u_0 +u_1|\chi|^2 +u_2|\chi|^4 +v_0 \chi^4 +v_1\chi^{*4} +O(\chi^6),
\end{equation}
where \(u_0,\:u_1,\:u_1,\:v_0\) and \(v_1\) are constants.

This potential presents the most general form of a \(C_4\) invariant potential close to the symmetry axis. Since we assume that the first medium is \(O(2)\) invariant, it is clear that the only terms that break the symmetry into \(C_4\) are \(\chi^4\) and \(\chi^{*4}\). Since we are only analyzing the result of the symmetry breaking process, it is sufficient to only consider the \(\chi^4\) and \(\chi^{*4}\) terms. We take \(u_0 =u_1 =u_2=0\) and proceed to evaluate the form of the function after the action of the symmetry breaking impulse. 

By only considering the symmetry breaking terms, our evolution operator becomes
\begin{equation}\label{exptaylor}
e^{i \Delta V_0 \Delta\tau}\rightarrow e^{i \Delta\tau (v_0 \chi^4 +v_1 \chi^{*4})}.
\end{equation}


\section{Energy}
\label{sec:energy}

Let us evaluate the energy in the 3D formalism. The inner product of \({\bf r'}\) with \({\bf r''}\) is
\begin{align}\nonumber
{\bf r'}\cdot{\bf r''}&={\bf r}\cdot(-\Omega_0^2{\bf r}+({\bf r}\times\Lambda))\\\nonumber
&=-\Omega_0^2({\bf r'}\cdot{\bf r})+{\bf r'}\cdot({\bf r}\times \Lambda)\\\nonumber
&=-\Omega_0^2({\bf r'}\cdot{\bf r})+\Lambda\cdot({\bf r'}\times{\bf r})\\
&=-\Omega_0^2({\bf r'}\cdot{\bf r})-\Lambda\cdot{\bf L}.
\end{align}
We can rearrange for \(\Lambda\cdot{\bf L}\) to get
\begin{equation}\label{lambda}
-\Lambda\cdot{\bf L}={\bf r'}\cdot{\bf r''}+\Omega_0^2({\bf r'}\cdot{\bf r}).
\end{equation}
The definition of energy is 
\begin{align}
E&=\frac{1}{2}({\bf r'}\cdot{\bf r'})+\frac{1}{2}\Omega_0^2({\bf r}\cdot{\bf r}),\\\nonumber
\end{align}
and then
\begin{align}
\frac{dE}{d\tau}&=\frac{1}{2}\frac{d}{d\tau}({\bf r'}\cdot{\bf r'})+\frac{1}{2}\frac{d}{d\tau}(\Omega_0^2{\bf r}\cdot{\bf r})\\\nonumber
&=\frac{1}{2}({\bf r''}\cdot{\bf r'}+{\bf r'}\cdot{\bf r''})+\frac{1}{2}\frac{d}{d\tau}(\Omega_0^2{\bf r}\cdot{\bf r})\\\nonumber
&={\bf r'}\cdot{\bf r''}+\frac{1}{2}\left(\frac{d\Omega_0^2}{d\tau}({\bf r}\cdot{\bf r})+\Omega_0^2({\bf r'}\cdot{\bf r}+{\bf r}\cdot{\bf r'})\right)\\
&={\bf r'}\cdot{\bf r''}+\Omega_0^2({\bf r}\cdot{\bf r'})+\frac{1}{2}\frac{d\Omega_0^2}{d\tau}
\end{align}
We use \eqref{lambda} and the expression for \(\Omega_0^2\) to obtain
\begin{equation}
\frac{dE}{d\tau}=-\Lambda\cdot{\bf L}+\frac{1}{2}\frac{d\Omega_0^2}{d\tau}=-\Lambda\cdot{\bf L}+\frac{3}{4\tau^3}.
\end{equation}
Finally, if we use our value for \(\Lambda\cdot{\bf L}\), we see that the change in energy is consistent with our previous analysis in \eqref{energy}
\begin{equation}
\frac{dE}{d\tau}=\frac{1}{2}\left(|{\bf r}|^2\Omega_1^2+\frac{3}{4\tau^3}\right)=\frac{3}{4\tau}\left(\frac{1}{\tau^2}+\frac{1}{4}\right).
\end{equation}

\end{document}